\documentclass[twocolumn,aps,prl,draft,showpacs]{revtex4}
\usepackage{graphicx}
\usepackage{bm}
\begin{document}
\def\be{\begin{equation}}
\def\ee{\end{equation}}
\def\bea{\begin{eqnarray}}
\def\eea{\end{eqnarray}}
\def\E{{\rm e}}
\def\bearst{\begin{eqnarray*}}
\def\eearst{\end{eqnarray*}}
\def\peleven{\parbox{11cm}}
\def\peffec{\peight{\bearst\eearst}\hfill\peleven}
\def\pspace{\peight{\bearst\eearst}\hfill}
\def\ptwelve{\parbox{12cm}}
\def\peight{\parbox{8mm}}

\title{Relation between shear parameter and Reynolds number in statistically 
stationary turbulent shear flows}
\author{J\"org~Schumacher}
\affiliation{Fachbereich Physik, Philipps-Universit\"at, 
             D-35032 Marburg, Germany}
\date{\today}            
\begin{abstract}
Studies of the relation between the shear parameter $S^*$ and the
Reynolds number $Re$ are presented for a nearly homogeneous and
statistically stationary turbulent shear flow.  The parametric
investigations are in line with a generalized perspective on the return
to local isotropy in shear flows that was outlined recently [Schumacher,
Sreenivasan and Yeung, Phys.  Fluids {\bf 15}, 84 (2003)].  Therefore,
two parameters, the constant shear rate $S$ and the level of initial
turbulent fluctuations as prescribed by an energy injection rate
$\epsilon_{in}$, are varied systematically.  The investigations suggest
that the shear parameter levels off for larger Reynolds numbers
which is supported by dimensional arguments.  It is found that the
skewness of the transverse derivative shows a different decay behavior with
respect to Reynolds number when the sequence of simulation runs follows
different pathways across the two-parameter plane.  
The study can shed new
light on different interpretations of the decay of odd order moments in
high-Reynolds number experiments.
\end{abstract}
\maketitle
\section{Introduction}
Homogeneous shear turbulence can be considered as the first non-trivial
extension of homogeneous, locally isotropic turbulence.  It is a flow that 
is thought as a bridge between the strongly idealized homogeneous isotropic
turbulence and more realistic turbulent shear flows such as channel flows
\cite{pope2000}.  A constant mean shear rate $S$ is present which is 
a large-scale source of anisotropy and the question that immediately arises
is how the statistical properties at the smallest scales of the turbulent flow
are affected by its presence.  After the pioneering work of Lumley
\cite{lumley1967} in which he predicted by dimensional arguments a
rather rapid decay of such anisotropy, namely with $\sim R_{\lambda}^{-1}$, this
particular flow came under renewed interest in the last decade. 
Several systematic measurements in simple shear flows
\cite{gargwarhaft1998,ferchichitavoularis2000,shenwarhaft2000,staicuvandewater2003} 
and in atmospheric boundary layers for the largest
accessible Taylor microscale Reynolds numbers of $R_{\lambda}\sim 10^4$
were presented.\cite{kuriensreenivasan2000}  All experiments detected
systematic deviations from Kolmogorov's concept of local isotropy.
\cite{kolmogorov1941} The decay of odd order normalized transverse
derivative moments with respect to Taylor microscale Reynolds number occurs
with a larger exponent than -1
and the
co-spectrum of the shear stress deviates from the classical
-7/3 law.\cite{kuriensreenivasan2000}  For
moderate Reynolds numbers, such a persistence in the decay of odd 
order moments was
also found within direct numerical simulations (DNS) 
and its connection to coherent structures and intermittency corrections 
could be
addressed.\cite{pumir1996,schumachereckhardt2000,schumacher2001,gualtierietal2002,ishiharaetal2002}

Recently, the record of experimental and DNS data on homogeneous and nearly
homogeneous shear flows was collected and discussed anew from a generalized
perspective by taking into account the role of small-scale intermittency and
mean shear.\cite{schumacheretal2003}  It was found that the operating points of
all those experiments are scattered in a two-parameter plane which is spanned by
the Reynolds number $Re$ and the shear parameter $S^*$.  Different experiments
followed different pathways across such plane causing, e.g., variations in the
decay behavior of odd order moments when simply projected onto the Reynolds
number axis as it is done usually when the issue of isotropy is discussed.  This
was identified as one possible reason for different interpretations of the data
in terms of the return to local isotropy for higher Reynolds numbers.

\begin{figure}
\centerline{\includegraphics[angle=0,scale=0.5,draft=false]{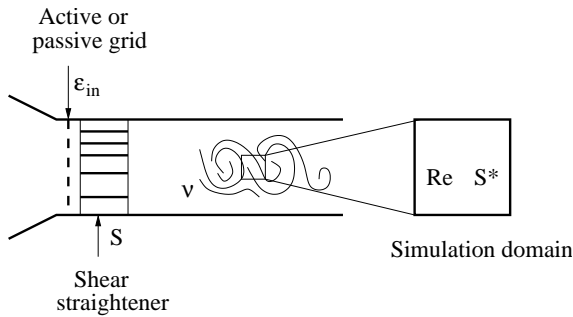}}
\caption{Sketch of a typical experimental situation in a simple shear flow
and its relation to the present investigation. A shear straightener sets
up a mean flow with a shear rate $S$ and an active or passive grid produces
additional turbulent fluctuations. The turbulence is decaying along
the downstream direction. The small volume would be the one we 
model within the DNS.}  
\label{fig1}
\end{figure}

One outcome of Ref.~14 is the necessity for more systematic numerical experiments
which will be presented in the following.  Here, two system parameters that
determine the homogeneous shear flow will be varied:  the constant shear rate,
$S$, and an energy injection rate, $\epsilon_{in}$, that prescribes the amount
of initially isotropic turbulent fluctuations.  The latter parameter can also be
thought of as a substitute for active or passive grids that are often placed in
wind tunnel experiments before the working fluid enters the shear
straightener (see e.g. Refs. 4 to 6 and the
sketch in Fig.~\ref{fig1}).  Such additional device (which is optional, but
frequently used) increases the turbulent fluctuations and thus the Reynolds
number while operating with the same mean shear rate $S$.  Beside this main
motivation coming from the experiments in nearly homogeneous shear flows, it
allows here for reaching a stable turbulent regime for small shear rates.
Consequently, the numerical experiments will give us hints on the following
functional dependencies
\begin{eqnarray}
\label{relation1}
Re&=&F_1(\nu,S,\epsilon_{in})\,,\\
S^*&=&F_2(\nu,S,\epsilon_{in})\,,
\label{relation2}
\end{eqnarray}
and would therefore allow for a more systematic study of the deviations from
local isotropy, e.g.  how transverse derivative moments decay along particular
pathways across the two-parameter plane.  We can simplify the functional
dependencies in relations (\ref{relation1}) and (\ref{relation2}) by keeping the
kinematic viscosity $\nu$ constant throughout the study.

Numerical models for homogeneous shear turbulence can be divided into three
different groups.  The first one uses finite difference methods with shear
periodic boundary conditions on the grid in physical space.\cite{gerzetal1989}
The other two approaches use pseudospectral techniques with periodic boundary
conditions.  Rogallo suggested a time-dependent coordinate transformation for
the inclusion of constant shear in such a simulation of turbulence.
\cite{rogallo1981,rogersmoin1986} The consequence is that the numerical grid
gets steadily skewed and has to be remeshed after time $t=2/S$.  In this method,
it was hard to reach a stationary turbulent state, the integral length scale
grows exponentially in time until the order of the box length is reached.
Statistical investigations that continued afterwards had to be run for very long
time intervals due to large fluctuations of energy or enstrophy.
\cite{pumir1996,gualtierietal2002,yakhot2003} In Ref.~9 integrations up to
$St\sim 10^3$ were performed.  For very large $S$ aliasing errors were
reported.\cite{leeetal1990} A method to overcome some of the problems was
suggested recently.\cite{schumachereckhardt2000}  The method avoids the
remeshing procedure.  A statistically stationary state with smaller fluctuations
of energy and enstrophy can be attained by an appropriate volume forcing in
combination with free-slip boundary conditions.  The boundary conditions will
cause small deviations from transverse homogeneity, so strictly speaking our
system is a nearly homogeneous shear flow, as it is the case for the
measurements.

The outline of the manuscript is as follows.  In the next section, we present in
brief the numerical model and the volume forcing schemes which are related to
both outer parameters, $S$ and $\epsilon_{in}$.  In section III we conduct
detailed studies of the statistical stationarity and point to differences in
comparison with the classical remeshing approach.  Afterwards, we are going to
discuss the relation between shear parameter and Reynolds number.  Predictions
that are made on the basis of dimensional arguments will be compared with
numerical findings.  Based on these results, the behavior of derivative moments
with respect to Reynolds number and shear parameter will be studied.  Finally, a
summary and an outlook are given.

\section{Numerical model}
\subsection{Equations and boundary conditions}

With length scales measured in units of the box width $L_y$, and time scales in
units of the eddy turnover time $L_y/v_{rms}$, the dimensionless form of the
equations for an incompressible Navier--Stokes fluid become
\begin{eqnarray}
\label{nseq}
\frac{\partial{\bf u}}{\partial t}+({\bf u}\cdot{\bf \nabla}){\bf u}
&=&-{\bf \nabla} p+\frac{1}{Re}{\bf \nabla}^2{\bf u}+{\bf F}\;,\\
\label{ceq}
{\bf \nabla}\cdot{\bf u}&=&0\;,
\end{eqnarray}
where $p({\bf x},t)$ is the pressure, ${\bf u}({\bf x},t)$ the velocity field.
The large scale Reynolds number is then 
\begin{eqnarray}
Re=\frac{v_{rms} L_y}{\nu}\,.
\label{reynoldsnumber}
\end{eqnarray}
The velocity components are decomposed in a mean fraction and a fluctuating 
turbulent part, $u_i=\langle u_i\rangle+v_i$ for $i=x, y$ and $z$, the
so-called Reynolds decomposition.  
The mean flow profiles are given for the homogeneous shear flow by
\begin{equation} 
\label{meanprofiles}
\langle u_x\rangle_{A,T}=S y,\,\,\,\langle u_y\rangle_{A,T}=\langle
u_z\rangle_{A,T}=0, 
\end{equation} 
where $x$ are streamwise (or downstream), $y$ shear (or wall-normal), and $z$
spanwise directions, respectively.  
Statistical averages are denoted by parentheses for the following and indices
indicate whether a time average, $\langle\cdot\rangle_T$, a volume average,
$\langle\cdot\rangle_V$, an average over $x-z$ planes at fixed y, 
$\langle\cdot\rangle_A$, or combinations of them are taken. Consequently,
the root mean square velocity reads 
$v_{rms}=\sqrt{\langle {\bf v}^2\rangle_{V,T}}$ and
the dimensionless shear parameter becomes 
\begin{eqnarray} 
S^*=\frac{S v^2_{rms}}{\epsilon}\,,
\label{shearparameter}
\end{eqnarray}
with the energy dissipation rate 
$\epsilon=\langle\epsilon({\bf x},t)\rangle_{V,T}$ and 
$\epsilon({\bf x},t)=(\nu/2)\sum_{i,j}(\partial_j v_i({\bf x},t)+
\partial_i v_j({\bf x},t))^2$. 

The pseudospectral method is applied.  The equations are integrated by a second
order predictor-corrector scheme where the time stepping satisfies the
Courant-Friedrichs-Levy criterion.  The Courant number was always below 1/2.
The dissipative term was included as an integrating factor $\exp(-\nu k^2t)$ for
every mode ${\bf u}_{\bf k}(t)$.  De-aliasing is done by a combination of
2/3-rule truncation and phase-shifting \cite{rogallo1981}.  As a criterion for
sufficient spectral resolution $k_{max} \eta > 1$ is used \cite{pope2000} with
Kolmogorov length scale $\eta=(\nu^3/\epsilon)^{1/4}$ and $k_{max}=\sqrt{2}
N/3$.  The aspect ratio $L_x:L_y:L_z=2\pi:\pi:2\pi$ was resolved with
$N\times(N/2+1)\times N$ grid points where $N=256$ and 512, respectively .

We take periodic boundary conditions in streamwise and spanwise directions,
respectively. In the shear direction, free-slip boundary conditions are applied,
\begin{equation} 
\label{bc}
u_y=0,\;\;\text{and}\;\;\frac{\partial u_x}{\partial y}=
\frac{\partial u_z}{\partial y}=0.
\end{equation} 
The free-slip boundary conditions will cause slight deviations from the
transverse homogeneity, an effect which decreases with growing Reynolds 
number \cite{eckhardtschumacher01}.
In contrast to the no-slip case, no energy is transfered into the flow
via the free-slip boundaries and thus an additional volume forcing ${\bf F}$
has to be applied (cf. eq.~(\ref{nseq})) in order to sustain turbulence. 

\subsection{Forcing scheme}
A compact formulation of the Navier-Stokes equations in Fourier space for our 
problem is given by
\begin{eqnarray}
\label{nseqF}
\frac{\partial u_{i{\bf k}}}{\partial t}
&=&N_{i{\bf k}}(u_{j{\bf q}}, t; Re)+F_{i{\bf k}}(t)\;,
\end{eqnarray}
where $N_{i{\bf k}}$ contains all nonlinear mode coupling contributions and the
dissipative part, $-Re^{-1} k^2 u_{i{\bf k}}(t)$.  In the following we will
combine two kinds of forcing, a volume forcing that sustains the linear mean
flow profile which will be referred to as {\em shear forcing} and a
homogeneous forcing that mimics the injection of additional fluctuations by
the grid which will be referred to as {\em grid forcing}.

A handful of Fourier modes is driven now in order to give an almost linear
profile for $\langle u_x\rangle_{A,T}$, i.e.  the forcing is with respect to the
streamwise velocity component only.  The driven modes are for the wave vectors
${\bf k}^{\prime}=(0,2n+1,0)$ for $n=0$ to 5.  They form the almost linear
mean profile,
\begin{eqnarray}
\label{mean}
\langle u_x\rangle_{A,T}(y)=
\frac{2\alpha}{\pi}\left(y-\frac{\pi}{2}\right)
\approx-\frac{8\alpha}{\pi^2}\sum^5_{n=0}
\frac{\cos[(2n+1)y]}{(2n+1)^2}\,,
\end{eqnarray}
for $y\in[0,\pi]$. This results in a mean shear rate $S=2\alpha/\pi$. 
The shear forcing is chosen in such a way that just these modes (which have 
real parts 
only for symmetry reasons) are held fixed to the amplitude 
values as given by Eq.~(\ref{mean}), i.e.
\begin{eqnarray}
\label{forcing1}
\frac{\partial u_{x{\bf k}^{\prime}}}{\partial t}
=0\;\;\;\;\Leftrightarrow\;\;\;\;
F^{(s)}_{x{\bf k}^{\prime}}(t)=-N_{x{\bf k}^{\prime}}(u_{j{\bf q}}, t; Re)\;.
\end{eqnarray}
These coefficients are set at the begining and do not violate the boundary
conditions and the divergence-zero condition because  
the driven modes $u_{x{\bf k}^{\prime}}(t)$ are functions of $k_y^{\prime}$ only.
The particular kind of forcing in the Fourier space might be nonanalytic in
the real space and will vary in space and time.

Second, the grid forcing at small wavenumbers is applied that injects
a certain amount of energy per time unit into the flow given by the 
rate $\epsilon_{in}$, \cite{grossmannlohse1994}
\begin{eqnarray}
\label{forcing2}
{\bf F}^{(g)}_{{\bf k}^{\prime\prime}}(t)=\epsilon_{in}\frac{{\bf u}_{\bf k}(t)}
{\sum_{{\bf k}^{\prime\prime}\in K}|{\bf u}_{{\bf k}^{\prime\prime}}(t)|^2}\,
\delta_{{\bf k,k}^{\prime\prime}}\;.
\end{eqnarray}
Consequently, it would follow 
\begin{eqnarray}
\label{forcing3}
\sum_{\bf k} {\bf u}^*_{\bf k}(t)\cdot {\bf F}^{(g)}_{\bf k}(t)=
\epsilon_{in}\equiv\epsilon\,,
\end{eqnarray}
for the statistically stationary balance of the turbulent kinetic 
energy when $S\equiv 0$.
The wavenumbers of the set $K$ do not coincide with the 
ones for the shear forcing, but are also small. 

\section{Statistical stationarity}
\subsection{Difference to the remeshing method}
The Reynolds decomposition of eqns.~(\ref{nseq}) and 
(\ref{ceq}) for all fields
results in
\begin{eqnarray}
\label{reeq}
\langle u_j\rangle\partial_j\langle u_i\rangle
&=&-\partial_i\langle p\rangle+\frac{1}{Re}\partial_j^2\langle u_i\rangle+
\langle F_i\rangle-\partial_j\langle v_i v_j\rangle\;,
\end{eqnarray}
where we considered already the time independence of the mean profile. 
At this point, a difference to the case with remeshing becomes obvious. 
All terms of the equation will be zero independently of each other due to
exact homogeneity for the latter case.
The violation of the transverse homogeneity will cause relics in the Reynolds
balance for the present model and results in 
\begin{eqnarray}
\label{reeq1}
0=\langle F_x^{(s)}\rangle_{A,T}-\partial_y\langle v_x v_y\rangle_{A,T}\;.
\end{eqnarray}
\begin{figure}
\centerline{\includegraphics[angle=0,scale=0.5,draft=false]{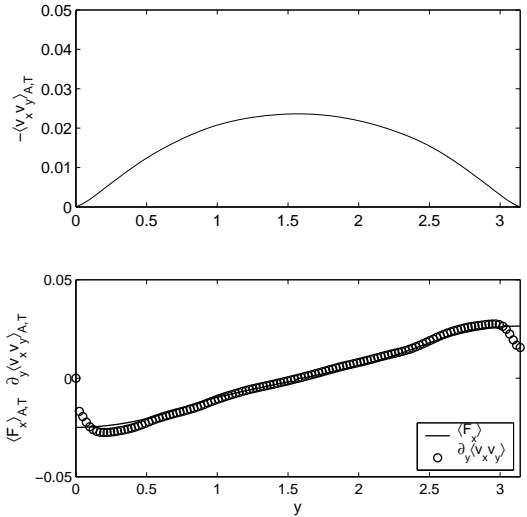}}
\caption{Upper panel:  
Reynolds shear stress $-\langle v_x v_y\rangle_{A,T}$ as a function of $y$.
Average was taken with respect to time and over $x-z$ planes at
fixed $y$. Lower panel: Verification of relation (\ref{reeq1}). It can 
be seen that
$\langle F_x^{(s)}\rangle_{A,T}$ and 
$\partial_y\langle v_x v_y\rangle_{A,T}$ collapse for
almost all $y$.}
\label{vpeps}
\end{figure}
Figure~\ref{vpeps} illustrates this behavior. We show the vertical profile 
of the shear stress, $\langle v_x v_y\rangle_{A,T}$ 
(upper panel) and verify that Eq.~(\ref{reeq1}) is 
satisfied to a good approximation except very close to the boundaries (lower
panel). The upper panel shows also that our system becomes nearly 
homogeneous only with respect to the transverse direction.
To conclude, 
there is a difference between both methods for the large scale balance.
The question is now: how are small-scale statistical 
properties of the fluctuating quantities affected in the present approach? 
As demonstrated in Refs.~10 and 11 small
scale statistical properties such as Reynolds stress magnitudes or higher
order normalized moments of the transverse derivative and spanwise vorticity, 
respectively, agree with previous homogeneous shear flow
simulations \cite{rogersmoin1986,pumir1996}
and experiments for the lowest Reynolds numbers \cite{shenwarhaft2000}.

\begin{figure}
\centerline{\includegraphics[angle=0,scale=0.5,draft=false]{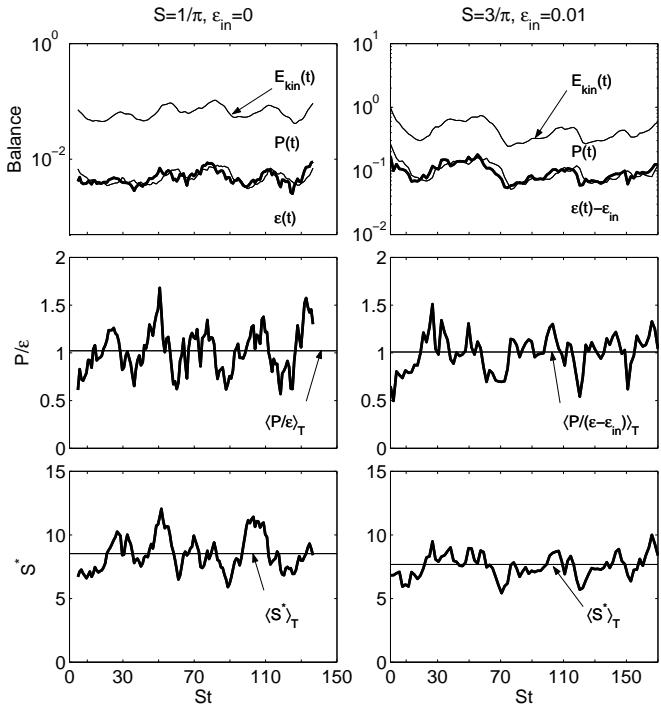}}
\caption{Statistical stationarity of the shear flow. Left column is for
run IV, the right column is for run IIc (see Tab.1). 
Upper panels: Temporal evolution of turbulent kinetic energy 
$E_{kin}(t)$, 
energy dissipation rate $\epsilon(t)$ (thin solid line), and production of turbulent
kinetic energy $P(t)$ (thick solid line). Mid panels: Temporal evolution of the
ratio $P(t)/\epsilon(t)$. The thin solid line is the temporal mean of the 
data and very close to one. Lower panels show the temporal variation of the shear parameter
$S^*$. Temporal means are again indicated by thin solid lines.}
\label{fig2}
\end{figure}
The forcing keeps the flow in a statistically stationary state with moderate
fluctuations of the kinetic energy and enstrophy, respectively (see
Ref.~11 for a more detailed investigation). 
In order to demonstrate this property of the current numerical scheme,
we show time series of the turbulent kinetic
energy, $E_{kin}(t)=q^2(t)/2=\langle {\bf v}^2\rangle_V/2$, the production of turbulent
kinetic energy, $P(t)=-\langle v_x v_y\rangle_V S$, and the energy dissipation
rate, $\epsilon(t)=\nu \langle (\partial_i v_j)^2\rangle_V$ for two runs
in Fig.~\ref{fig2} (run IIc and IV of Tab.~1). The fluctuations of the kinetic 
energy are by about a factor of 2 smaller than those reported 
in the long-time remeshing runs.\cite{pumir1996,gualtierietal2002} With
$\sigma_E=\sqrt{\langle(E_{kin}(t)-\langle E_{kin}\rangle_T)^2\rangle_T}$ we get
$\sigma_E/\langle E_{kin}\rangle_T=24\%$ for run IV and 33\% for run IIc.
This ratio decreases even further for runs Ia to Ie.
The resulting global
energy balance for the fluctuating part is then
\begin{eqnarray}
\label{tkeeq2}
\partial_t \frac{q^2}{2} &\approx&-\epsilon(t)+P(t)+\epsilon_{in}\,.
\end{eqnarray}
It was verified that contributions from shear driving are 
subdominant as well as the
transverse flux contributions from third order terms arising in the balance.
The reduced fluctuations of the kinetic energy might be caused due to the 
free-slip boundaries. As mentioned above, small boundary layers do form that 
cause slight transverse fluxes into the bulk.\cite{hill2001,casciolaetal2003}   
\begin{figure}
\centerline{\includegraphics[angle=0,scale=0.5,draft=false]{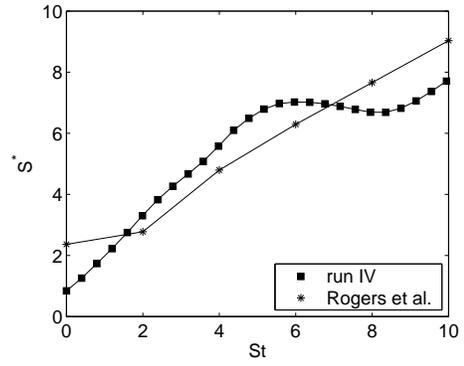}}
\caption{Temporal evolution of the dimensionless shear parameter in the 
initial phase of the simulation for run IV. The run started with  
isotropic turbulence decaying with a $k^{-5/3}$ spectral law and was 
asserted to the constant shear with rate $S$. For comparison we 
added initial data from Rogers {\it et al.} \cite{rogersetal86a} (run C128V).}
\label{fig3}
\end{figure}

As can be seen in  Fig.~\ref{fig2}, the ratio of $P/\epsilon$
varies in time, but its temporal mean is almost exactly unity and 
\begin{eqnarray}
\langle P(t)\rangle_T=\langle \epsilon(t)\rangle_T-\epsilon_{in}\,,
\end{eqnarray}
holds. We note that this property differs from the remeshing simulations
where the ratio $P/\epsilon$ is constant but larger unity. Quantities such
as the 
turbulent kinetic energy will grow then exponentially with time which is a 
consequence of (\ref{tkeeq2}), i.e. 
$q^2(t)\sim \exp[(P/\epsilon-1)t]$. One might conclude that the
violations of the transverse homogeneity cause this ratio to vary around unit 
value and thus to assure statistical stationarity.  This point should not be
mixed with the exclusion of statistical stationarity and downstream homogeneity
which is relevant in experiments \cite{harrisetal1977,tavoularis1985} 
but not in DNS. 
  
\begin{figure}
\centerline{\includegraphics[angle=0,scale=1.2,draft=false]{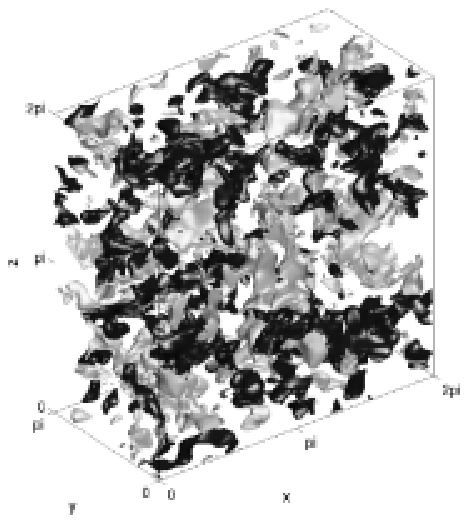}}
\centerline{\includegraphics[angle=0,scale=1.2,draft=false]{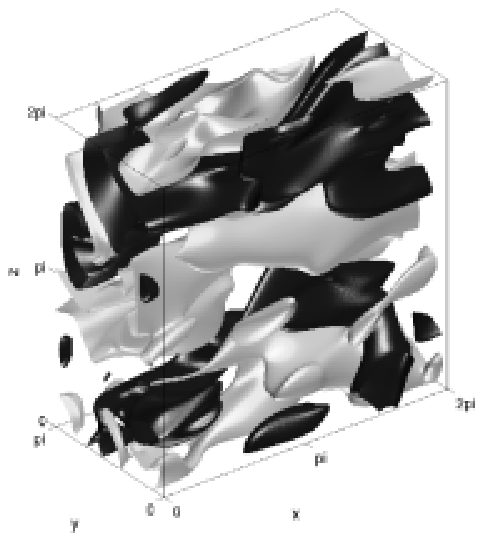}}
\caption{Isosurface plots of the streamwise turbulent velocity for the
initial phase of run IV. Level sets with opposite sign are shown in each
picture (grey is positive and black is negative). 
Upper panel: St=0.4. Lower panel: St=9.6.}
\label{fig4}
\end{figure}

It is interesting to take a closer look at the initial phase of
the evolution.  In Fig.~\ref{fig3} we plotted the initial evolution of the shear
parameter $S^*$ and for comparison we added DNS data by Rogers {\it et al.}
\cite{rogersetal86a}.  Both simulations show that for the initial phase an
(almost) linear growth can be detected.  It clearly illustrates the non-normal
amplification mechanism acting on the isotropic turbulent ``background''.
The streamwise velocity fluctuations grow due to the lift-up of
streamwise streaks, but will leave the energy dissipation rate which is dominant
at small scales close to its initial magnitude for a while.  In Fig.~\ref{fig4}
we show the streamwise turbulent velocity at two instants, an initial snapshot
(upper panel) where the fluctuations are still isotropic and a later snapshot 
(lower panel) where streamwise streaks have formed. A more detailed analysis
of the regeneration cycle of these streamwise streaks and vortices 
\cite{waleffe1997} in a shear flow with free-slip boundaries was discussed in
Ref.~28 and was found to agree qualitatively with other shear flows. 

\subsection{Integral scale}
\begin{figure}
\centerline{\includegraphics[angle=0,scale=0.5,draft=false]{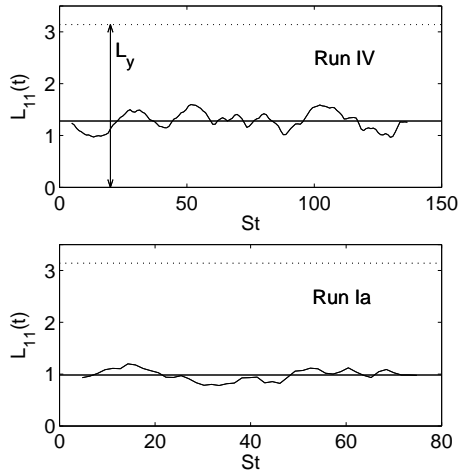}}
\caption{Temporal evolution of the integral length scale $L_{11}(t)$.
For comparison, the width of the simulation domain in shear direction
$y$ is indicated by the vertical double-headed arrow. 
The temporal average of $L_{11}(t)$ is
plotted as a solid line. Data are for run IV (upper panel) and run Ia
(lower panel).}
\label{fig5}
\end{figure}

In Fig.~\ref{fig5} we show the integral length scale which is defined 
as \cite{pope2000} 
\begin{eqnarray}
L_{11}(t)=\frac{\pi}{2\langle v_x^2\rangle}\,\int_0^{\infty}\,
\frac{E(k,t)}{k}\,dk\,,
\label{intlength}
\end{eqnarray}
with the energy spectrum, $E(k,t)$. In comparison to the simulations that use
the remeshing method of Rogallo, it can be seen that $L_{11}(t)$ remains well 
below the box width for our method and does not grow exponentially in time.    
It could be shown recently that strong fluctuations of the kinetic energy and
the enstrophy, respectively, are related to $L_{11}(t)$ reaching the size of 
the box length \cite{yakhot2003}. The present studies give $L_{11}(t)$ at
about 33\% of $L_y=\pi$ and are smaller as the reported 72\% to 80\% of
$L=2\pi$.\cite{pumir1996} 

\section{Investigations in the two-parameter plane}
\subsection{Estimates for the behavior of $S^*$ versus $Re$}
We turn now to investigations in the parameter plane that is spanned by the
two essential dimensionless parameters in a homogeneous shear flow, the shear
parameter $S^*$ and the Reynolds number $Re$. Every point $(S^*,Re)$
in this plane
will be denoted as an operating point of the particular 
simulation or 
measurement, it will drift across the plane for the non-stationary case. 
As was demonstrated in Ref.~14, the 
collected data were rather scattered over such a plane. 
Two limiting regimes for the homogeneous shear flow could be identified. 
One is the {\em large shear limit} which will approach the 
(linear) rapid distortion case for $S\to \infty$. A condition for
the large shear case follows from starting with
$| v_i \partial \langle u_j\rangle/\partial x_i| \gg | v_i \partial v_j/\partial x_i|$, 
and it is given by 
\begin{eqnarray}
S^* Re^{-1/2} \gg 1\,,
\label{largeshear}
\end{eqnarray}
in terms of the large scale Reynolds number $Re$. The second 
case is the {\em local isotropy limit} with the condition of
sufficient separation of large and small time scales $S\tau_{\eta}\ll 1$. 
$\tau_{\eta}=\sqrt{\nu/\epsilon}$ is the Kolmogorov time and it follows 
\begin{eqnarray}
S^* Re^{-1/2} \ll 1\,.
\label{localiso}
\end{eqnarray}
The idea 
is to identify a more systematic dependence $S^*(Re)$
in the plane, similar to a phase diagram in thermodynamics. It is clear
from the beginning that the range of parameters that we can cover with the DNS
experiments will be limited, but at least in a certain range we can 
conduct systematic studies.   

Before doing so, we  give simple arguments of which functional 
dependencies (\ref{relation1}) and (\ref{relation2}) might appear. Dissipation of energy is due
to shear effects and due to grid forcing and thus we can set the following
relation
\begin{eqnarray}
\epsilon=C_1\nu \frac{v_{rms}^2}{L_y^2}
+C_2\frac{v_{rms}^3}{L_y}\,,
\label{diss1}
\end{eqnarray}
where $C_1$ and $C_2$ are dimensionless constants.
The two terms on the r.h.s. of (\ref{diss1}) model the crossover 
of the energy dissipation rate from a weakly turbulent state to fully 
developed turbulence which is present
for moderate Reynolds numbers.\cite{doeringetal2003} While 
$\nu v_{rms}^2/L_y^2$ determines the dissipation at smaller Reynolds  
numbers, the term $v_{rms}^3/L_y$ takes over for larger Reynolds numbers 
(for more details see also Ref.~30).    
The root mean square velocity will
be determined by both large scale driving mechanisms and thus 
\begin{eqnarray}
v_{rms}=C_3\left(L_y \epsilon_{in}+C_4 L_y^3 S^3\right)^{1/3}\,,
\label{diss2}
\end{eqnarray}
can be taken by dimensional arguments 
with yet unknown dimensionless constants $C_3$ and $C_4$, respectively. 
Equations (\ref{diss1}) and (\ref{diss2}) are inserted into the relations 
for the Reynolds number (\ref{reynoldsnumber}) and the shear parameter
(\ref{shearparameter}), respectively. For both
parameters follow functions of $S$ and $\epsilon_{in}$
in the presence of constants $\nu$ and $L_y$  
(cf. Eqns.~(\ref{relation1}) and (\ref{relation2})) 
\begin{eqnarray}
S^*&=&\frac{L_y^2 S}
{C_1 \nu +
C_2 C_3 L_y \left(L_y \epsilon_{in}+C_4 L_y^3 S^3\right)^{1/3}}\,,
\label{diss3a}\\
Re&=&\frac{C_3 L_y \left(L_y \epsilon_{in}+C_4 L_y^3 S^3\right)^{1/3}}{\nu}\,.
\label{diss3b}
\end{eqnarray}
One point becomes obvious immediately:  there seems to be a variety of pathways
to explore the
$S^*$--$Re$ plane and the operating points can be scattered over a large
fraction. In fact, that is what the collected data points showed. 
\cite{schumacheretal2003}  

For simplicity, we study the behavior along curves
$S$=const and $\epsilon_{in}$=const. The Reynolds number grows to infinity 
for both cases because one of the 
sources for an increase of turbulent fluctuations is always
present. It reduces to
\begin{eqnarray}
Re_0=\frac{C_3 L_y^{4/3} \epsilon_{in}^{1/3}}{\nu}\,,
\end{eqnarray}
for turbulence without shear and marks the origin of curves $S^*(Re)$ in the
plane (cf. relations (\ref{largeshear}) and (\ref{localiso})).
We get the following limits for the shear parameter,
\begin{eqnarray}
\label{limit1}
& &\lim_{S=const, \epsilon_{in}\to\infty} S^* =0\,,\\
& &\lim_{\epsilon_{in}=const, S\to\infty} S^* =S_{\infty}^*=
\frac{1}{C_2 C_3 C_4^{1/3}}\,.
\label{limit2}
\end{eqnarray}
In limit
(\ref{limit1}) we observe that the shear parameter goes to zero with a power
${\cal O}(\epsilon_{in}^{-1/3})$. Physically, this means that the grid
driving becomes more and more important while shear effects decrease, i.e.  the
system approaches the isotropic limit $S^* (Re-Re_0)^{-1/2}\ll 1$,
which is the case to a good approximation in grid
turbulence. \cite{schumacheretal2003}  
On the other hand, when keeping $\epsilon_{in}$ fixed but
increasing external shear rate $S$, the estimate yields a non-zero
constant value if there is no further Reynolds number
dependence hidden in the four prefactors $C_i$.
This means that for a statistically stationary and nearly homogeneous system an
asymptotic value of $S^*$ is found. 

\subsection{Simulation results}
The outlined estimates are tested by numerical experiments. 
In Table~I, the different  parameter sets and the quantities are
summarized that are necessary for the calculation of $S^*$ and $Re$.
The DNS runs are ordered in three different series, No.~I varies along the 
isoline  $(S=const,\epsilon_{in})$, while series~II and III run 
along the iso-parameter line $(S, \epsilon_{in}=const.)$. Such investigation 
becomes rather expensive because every single data point in the parameter plane
is one long-time DNS run for a statistically stationary shear flow. 
\begin{table}
\begin{center}
\begin{tabular}{lccccccccc}
Run No.&$S$&$\epsilon_{in}$&$v_{rms}$&$\epsilon$&$Re$&$S^*$&$k_{max}\eta$&$\nu$&$N$\\
\hline
Ia       &$1/\pi$&0.010&0.47&0.017& 444&4.2&4.64& 1/300 &256 \\
Ib       &$1/\pi$&0.025&0.55&0.032& 520&3.0&3.96& 1/300 & 256 \\
Ic       &$1/\pi$&0.040&0.66&0.051& 622&2.7&3.52& 1/300 & 256 \\
Id       &$1/\pi$&0.055&0.67&0.063& 629&2.2&3.34& 1/300 & 256 \\ 
Ie       &$1/\pi$&0.100&0.86&0.114& 810&2.1&2.88& 1/300 & 256 \\ \hline
IIa      & 0 &    0.010&0.34&0.010& 325&0.0&5.29&  1/300 &256 \\
IIb (=Ia)&$1/\pi$&0.010&0.47&0.017& 444&4.2&4.64& 1/300 & 256 \\
IIc      &$3/\pi$&0.010&0.94&0.111& 891&7.7&2.90& 1/300 & 256 \\
IId      &$6/\pi$&0.010&1.49&0.480&1404&8.8&2.01& 1/300 & 256 \\ 
IIe      &$8/\pi$&0.010&3.39&2.998&3197&9.8&2.54& 1/300 & 512 \\ \hline
IIIa     & 0 &    0.025&0.48&0.025& 457&0.0&4.21& 1/300 & 256 \\
IIIb (=Ib)&$1/\pi$&0.025&0.55&0.032& 520&3.0&3.96&1/300 &256 \\
IIIc     &$6/\pi$&0.025&1.31&0.393&1235&8.4&2.11&  1/300 &256 \\ \hline
IV       &$1/\pi$&0 & 0.36 & 0.005 &  452  & 8.2& 5.07&  1/400 &256    
\end{tabular}
\label{tab1}
\vspace{0.5cm}
\caption{Parameters of the numerical experiments.  For convinience we listed the
runs in three series indicating different pathways in the parameter plane.  Run
IV was an additional long-time run for investigations on statistical
stationarity.}
\end{center}
\end{table}

\begin{figure}
\centerline{\includegraphics[angle=0,scale=0.5,draft=false]{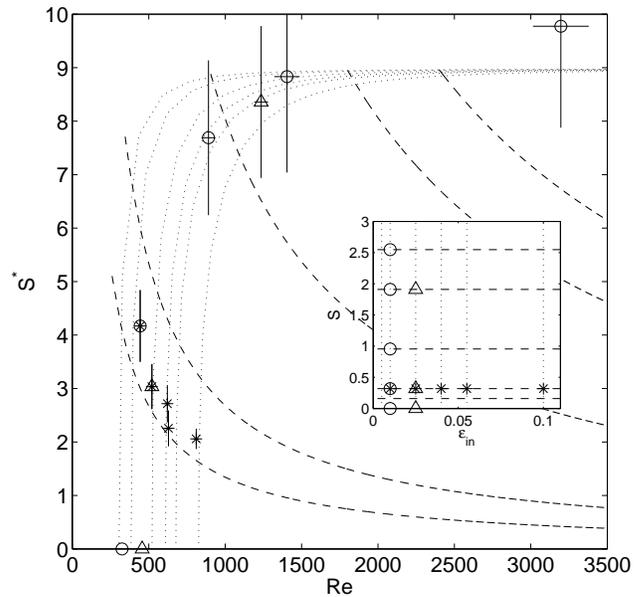}}
\caption{Operating points of statistically stationary homogeneous shear
flow simulations in the $Re$--$S^*$ parameter plane when iso-parameter 
lines were followed as indicated in the inset. Asterisks are data points
of series I, open circles of series II, and open triangles of series III.
The error bars 
($\pm\sigma/2$) are given in 
the panel. They follow from 
$\sigma_{S^*}=S^*(2\sigma_{v_{rms}}/v_{rms}+\sigma_{\epsilon}/\epsilon)$
and $\sigma_{Re}=Re\,\sigma_{v_{rms}}/v_{rms}$, respectively. 
The two data points with $S^*=0$ have error bars with respect to $Re$
which are about the size of the symbol. 
The mapping of the
isoline mesh of the inset to the $S^*$-$Re$ plane is also shown by
additional dotted and dashed lines lines. 
The constants are $C_1=1.5$, $C_2=0.35$, $C_3=1.29$,
and $C_4=0.015$.}
\label{fig6}
\end{figure}

In Fig.~\ref{fig6}, we have collected the operating points $(S^*, Re)$ of the
runs and added the error bars.  The inset illustrates the corresponding
variations in terms of the outer parameters.  We fitted the dimensional
estimates to the whole set of data points.  The dashed and the dotted 
lines are the fit results
of (\ref{diss3a}) and (\ref{diss3b}) to series I, II, and III.  The additional
lines illustrate how points $(S, \epsilon_{in})$ are mapped to $(S^*,
Re)$ with the fitted constants.  
It can be seen that the lines display the trends of the data
points quite well, but do not match perfectly with the data points.  Clearly,
the data base is rather sparse such that for a multi-dimensional fit 
problems can arise.  Second, we do observe that the error bars grow in magnitude
for the larger shear rates when the fluctuations of large scale velocity tend to
grow because the effect of non-normal streak lift-up gets more pronounced.

The numerical experiments support a saturation of the shear
parameter for larger Reynolds numbers which would mean that a flow with a given
injection rate can never cross the large shear limit boundary of 
$S^*(Re-Re_0)^{-1/2}\gg 1$ when S is increased.  Clearly, at the present 
stage this point
cannot be fully resolved, but we do observe a saturation along II and
III.  It might be the case that even for the present system the finite size of
the simulation domain affects the results eventually although the integral scale
remains well below the box size.  On the other hand and to our knowledge, all
DNS operated in a range of the shear parameter of $S^*\lesssim 10$
that could be exceeded only
transiently when the system was strongly non-stationary and therefore close to
the rapid distortion case \cite{rogersetal86a,leeetal1990}.

\begin{figure}
\centerline{\includegraphics[angle=0,scale=0.5,draft=false]{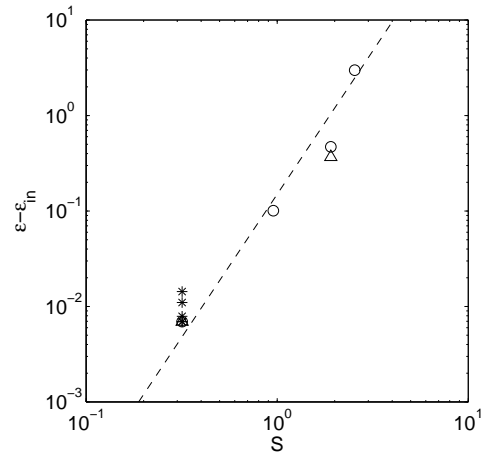}}
\caption{Double logarithmic plot of $\epsilon-\epsilon_{in}$ over $S$ for all 
three data sets. The symbols are the same as for Fig.~\ref{fig6}. The data
were fitted (least square) with a function $A S^2+B S^3$ where $A=10^{-4}$ and 
$B=0.15$. The dashed line shows $\epsilon-\epsilon_{in}=0.15 S^3$.}
\label{epsin}
\end{figure}

Further support for the ansatz (\ref{diss2}) is given by the following fact.
When inserting (\ref{diss2}) into (\ref{diss1}), one gets
$\epsilon\sim\epsilon_{in}+const\, S^3 $ to leading order plus further
subdominant terms in each of the variables.  In Fig.~\ref{epsin}, we plotted
$\epsilon-\epsilon_{in}$ over shear rate $S$ for all three series and observe
that for the larger shear rates a cubic power law fits quite well.  The scaling
is consistent with a direct proportionality between $S$ and 
$v_{rms}$ (cf.~(21)).

\begin{figure}
\centerline{\includegraphics[angle=0,scale=0.5,draft=false]{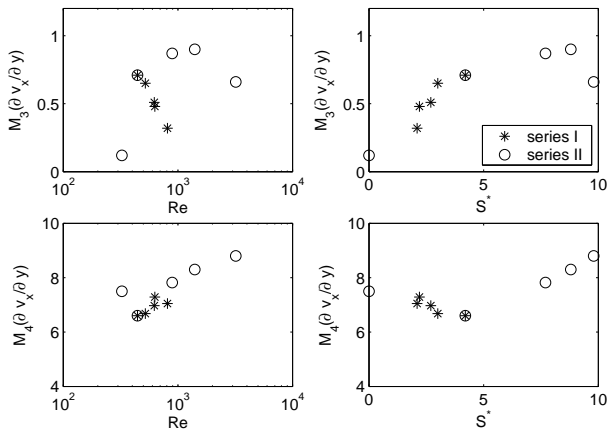}}
\caption{Normalized derivative moments $M_3(\partial v_x/\partial y)$ (upper
row) and $M_4(\partial v_x/\partial y)$ (lower row) as a function of the
Reynolds number $Re$ (left column) and the shear parameter $S^*$ (right column). 
The data from series I and II of Table~1 are plotted.}
\label{fig7}
\end{figure}

We turn back to the original scope of these investigations to get
a better understanding of the return to local isotropy in shear flows. 
We monitored derivative moments along the parameter curves. The normalized
$n$-th order moment of the transverse derivative of the streamwise turbulent
velocity component is given by
\begin{eqnarray}
M_{n}(\partial v_x/\partial y)= \frac{\langle (\partial
v_x/\partial y)^{n} \rangle}
     {\langle (\partial v_x/\partial y)^2 \rangle^{n/2}}\,.
\end{eqnarray}
The cases $n=3$ and 4 are studied and the results are 
summarized in Fig.~\ref{fig7}.
The flatness ($n=4$) in the lower row grows slowly with 
Reynolds number and shear parameter, respectively. The Reynolds number
trend is thus in agreement with previous investigations 
\cite{pumir1996,rogersmoin1986,schumachereckhardt2000,schumacher2001}.  
More interesting are the results for the third order moment, the skewness.
While it decreases strongly with $Re$ along path I it starts to decline
slowly only for the largest values of $Re$ or $S^*$ along path II. We note that 
for both series the Reynolds number is growing while their decay differs. 
A simple projection onto $Re$ gives thus
different results and underlines the idea of taking into account the 
shear rate as well. 
This is also supported when plotting all skewness data over $S^*$. 
Now, a trend of the data to have larger skewness values with growing 
shear parameter is observed 
(cf. upper right panel of Fig.~\ref{fig7}) for the majority of the data set. 

In Ref.~14 the data were discussed with respect to the following question:
Is a large Reynolds number really large if the shear parameter is large as well?
Large is meant there in the sense of being close to the local isotropy limit.
Based on our findings and on (\ref{limit2}) one can draw the conclusion that 
large means then  
\begin{eqnarray}
\label{limit3}
Re\gg (S_{\infty}^*)^2=\frac{1}{C_2^2 C_3^2 C_4^{2/3}}\,,
\end{eqnarray}
which can be further simplified.
For larger $Re$ one would expect $C_2\sim {\cal O}(1)$ because this constant 
is
nothing else but the dimensionless energy dissipation rate (see ansatz
(\ref{diss1})). A similar argument should hold for
$C_3$. Thus $Re\gg1/C_4^{2/3}$ would follow eventually. The remaining 
constant $C_4$ which relates the turbulent fluctuations to the shear rate 
cannot be evaluated by such arguments. The present DNS give $C_4\ll 1$ 
consistently (see
the caption of Fig.~\ref{fig6}). This can be an interesting point to be 
answered when the data base is more comprehensive.  

\section{Summary and outlook}
We have presented systematic studies of the relation between the shear parameter
and the Reynolds number in a turbulent nearly homogeneous shear flow.  The
present system allows for investigations of a statistically stationary flow with
mild fluctuations of the turbulent kinetic energy.  The ratio of the turbulence
production to the energy dissipation was found to vary around unity and the
integral length scale remains well below the box extension.  Two
parameters, the shear rate $S$ and the energy injection rate $\epsilon_{in}$,
were varied by keeping one constant in each case.  We monitored the resulting
shear parameter $S^*$ and Reynolds number $Re$.  It was found that they evolve
on different pathways across the two-parameter plane.
Dimensional estimates are in qualitative agreement with the results of the DNS.
They suggest that an asymptotic shear parameter value is reached for the
statistically stationary case.  The reason for such saturation might be due to
the finiteness of the considered volume, namely that the linear mean profile in
a homogeneous shear flow causes always an outer flow gradient scale that is 
beyond the box extensions (strictly speaking this scale is infinite). For any 
finite system the initially observed self-similar growth of quantities 
with time is interrupted once $L_{11}$ comes closer to the box size
and consequently the shear parameter settles to a finite 
value.  

Another way of such a two-parameter
discussion is to use the shear parameter $S^*$ and the Corrsin 
parameter $S_c=S\tau_{\eta}$ 
\footnote{This was pointed to us by one of the referees.} where the Taylor microscale Reynolds 
number follows to $R_{\lambda}\simeq S^*/S_c$. \cite{gualtierietal2002}
Clearly, our results can be transformed into the latter frame. The large 
shear limit from Ref.~14,
$S^* R_{\lambda}^{-1}\gg 1$, would thus translate to $S_c\gg 1$ and the local 
isotropy limit, $S^* R_{\lambda}^{-1}\ll 1$, follows to $S_c\ll 1$. The latter
inequality states then that the shear time scale $S^{-1}$ and the dissipative 
time scale $\tau_{\eta}$ have to be separated far enough of each other.    

It was shown that the third order derivative
moments decay with different trends with respect to the Reynolds number when
they are monitored along different pathways in the two-parameter plane.  There
is practically no decay over a wide range
when the same data are shown with respect to the the shear parameter.
All this might explain the different interpretations of measured moments in
terms of a return to local isotropy.

Clearly, the present the Reynolds numbers are mostly small and the data base is
rather sparse.  It has to become more comprehensive for some reasons:  first,
one would like to see of these trends persist to higher Reynolds numbers,
especially the behavior of the shear parameter $S^*$.  Secondly, the four
parameters $C_1$ to $C_4$ are assumed to be constant.  We do not know if a
Reynolds number dependence is hidden there; the current data did not allow for
drawing conclusions on that.  Thus, we consider this investigation as a starting
point.  Further extensions beside these two points are possible.  The so-called
grid forcing scheme gives us a tool into hands which can be extended to cases
where spatial and temporal patterns are used for the excitation of
turbulent fluctuations, an approach that recently started in terms of active
grids in wind tunnels.

\acknowledgments
The author would like to thank B.~Eckhardt, K.~R.~Sreenivasan and P.~K.~Yeung
for numerous discussions and helpful comments.  He is also grateful for the
hospitality and the financial support of the International Center for
Theoretical Physics in Trieste (Italy) where parts of the work were completed.
Financial support by the Deutsche Forschungsgemeinschaft is also acknowledged.  The
numerical simulations were carried out on a Cray SV1ex of the John von Neumann
Institute for Computing J\"ulich (Germany), on an IBM Power4 Regatta system at
the Hochschulrechenzentrum Darmstadt (Germany), and on the IBM
Blue Horizon at the San Diego Supercomputer Center within the NPACI initiative
of the National Science Foundation.

\end{document}